\documentclass{emulateapj}
\usepackage{graphicx}

\shorttitle{2M~0518, Discovery of an Unresolved L/T Binary}
\shortauthors{Cruz, Burgasser, Reid, \& Liebert}

\begin{document}

\title{2MASS~J05185995$-$2828372: Discovery of an Unresolved L/T Binary
}

\author{Kelle L. Cruz\altaffilmark{1,2}}

\author{Adam J. Burgasser\altaffilmark{2,3,4}}

\author{I. Neill Reid\altaffilmark{1,2,5}}

\author{James Liebert\altaffilmark{6}}

\altaffiltext{1}{Department of Physics and Astronomy, University of
Pennsylvania, 209 South 33rd Street, Philadelphia, PA 19104-6396;
kelle@sas.upenn.edu}

\altaffiltext{2}{Visiting Astronomer at the Infrared Telescope
Facility, which is operated by the University of Hawaii under
Cooperative Agreement no. NCC 5-538 with the National Aeronautics
and Space Administration, Office of Space Science, Planetary
Astronomy Program.}

\altaffiltext{3}{Division of Astronomy and Astrophysics, University of
California at Los Angeles, Los Angeles, CA, 90095-1562;
adam@astro.ucla.edu}

\altaffiltext{4}{Hubble Fellow}

\altaffiltext{5}{Space Telescope Science Institute, 3700 San
Martin Drive, Baltimore, MD 21218; inr@stsci.edu}

\altaffiltext{6}{Department of Astronomy and Steward Observatory,
University of Arizona, 933 North Cherry Avenue, Tucson, AZ 85721;
liebert@as.arizona.edu}

\begin{abstract}

We present the peculiar near-infrared spectrum of the newly discovered
brown dwarf 2MASS~J05185995$-$2828372, identified in the Two Micron All
Sky Survey.  Features characteristic of both L and T dwarfs are present,
namely strong carbon monoxide absorption in $K$-band, strong methane
absorption in $J$- and $H$-bands, and red near-infrared colors.  We
consider several scenarios that could produce these features and conclude
that the object is most likely to be an unresolved L/T binary system.  We
discuss how the estimated photometric properties of this object are
consistent with the observed $J$-band brightening of brown dwarfs between
late-L and early-T dwarfs, making detailed study of this system an
important probe of the L/T transition.

\end{abstract}


\section{Introduction}\label{sec:intro}

The existence of brown dwarfs, low-mass (M~$\la0.075$~M$_{\sun}$) objects
that form like stars but are incapable of maintaining core hydrogen
fusion, was first postulated by \citet{Kumar}. After several decades of
unsuccessful searches, well over 100 brown dwarfs are currently known,
primarily as a consequence of the availability of deep far-red and
infrared sky surveys \citep[DENIS, 2MASS,SDSS]{DENIS,2MASS,SDSS}.  Without
a long-lived energy source, brown dwarfs cool rapidly, exhibiting spectra
dominated by a sequence of complex molecular bands. Metal hydride
absorption (e.g., FeH, CrH, and CaH) replaces titanium oxide at optical
wavelengths as the effective temperature falls below 2100 K, and the
spectral type evolves from type M to type L \citep{K99,Chabrier}. As the
temperature drops below 1300 K, methane forms in the outer atmosphere
\citep{Tsuji64} and the strong absorption at 1--3~$\micron$ leads to
significantly bluer near-infrared colors. These are T dwarfs
\citep{B02,Geballe}.

We are currently using near-infrared photometry from the Two Micron
All-Sky Survey \citep[2MASS]{2MASS} to search for all late-type M and L
dwarfs  lying within 20 parsecs of the Sun \citep{paper5}. In the course
of follow-up observations, we have identified a cool dwarf which appears
to break the current spectral classification paradigm.
2MASS~J05185995$-$2828372 (hereafter, 2M~0518) was selected for
observation based on its red ($J-K_s$) color of 1.82 magnitudes and its
relatively bright apparent magnitude, $J=15.98$.  The peculiar
near-infrared spectrum of this object, however, exhibits both L and T
dwarf spectral features.  The following section describes our observations
and \S~\ref{sec:discussion} discusses possible explanations for the
observed properties of this intriguing object.

\begin{figure}[b]
\centering
\includegraphics[width=3.25in]{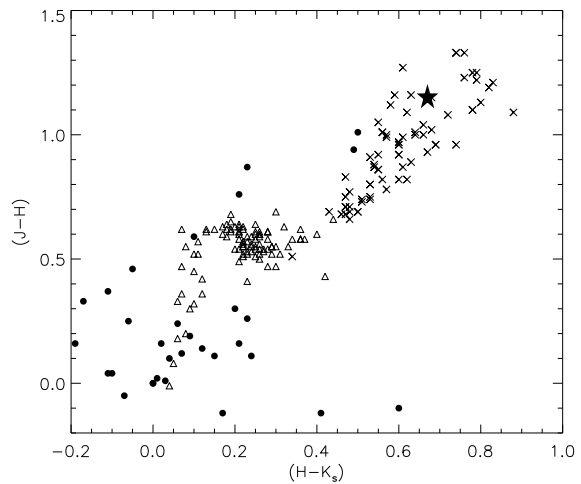}
\caption{Color-color diagram for 2M~0518 (\emph{five-pointed star}),
late-type stars (\emph{triangles}), L dwarfs (\emph{crosses}), and T
dwarfs (\emph{circles}).}\label{fig:color}
\end{figure}

\section{Observations}\label{sec:obs}

\begin{figure}[t]
\centering
\includegraphics[width=3in]{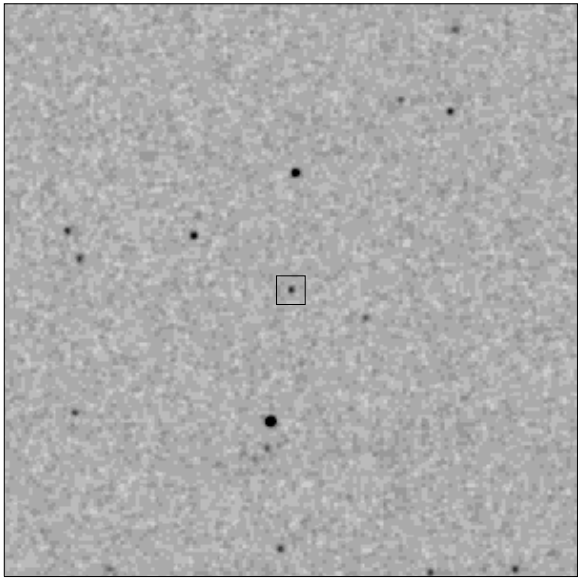}
\caption{$K_S$-band finder chart for 2M~0518 as taken from the 2MASS
All-Sky Quicklook Image Service. The epoch of the image is 1999 January
06. Images are 5\arcmin\ on each side, with north up and east to the
left.}\label{fig:finder}
\end{figure}

\begin{figure*}[t]
\centering
\includegraphics[width=4.5in]{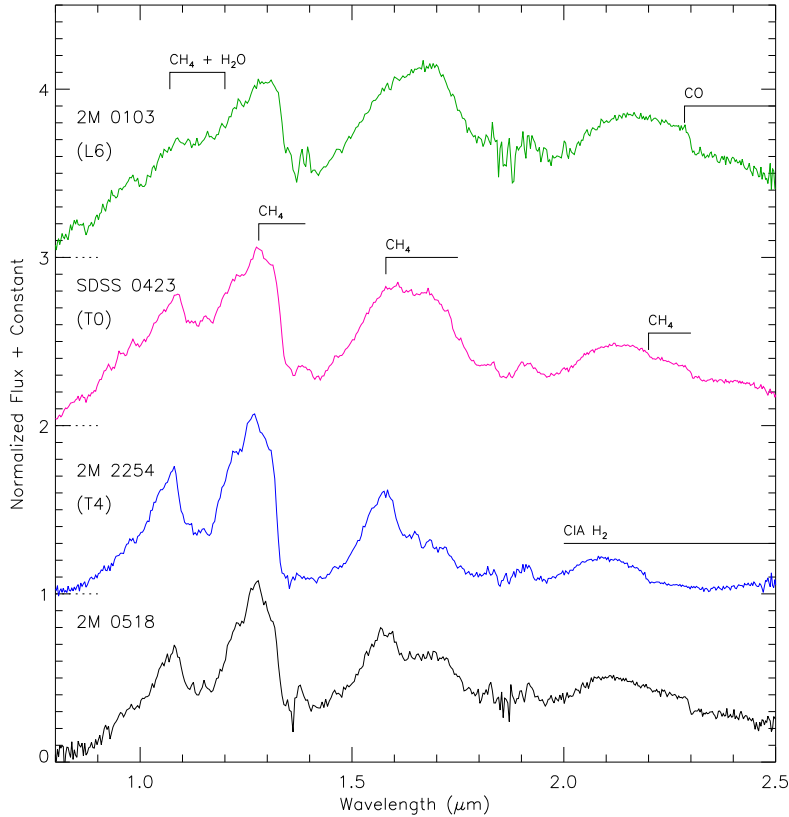}
\caption{Spectrum of 2M~0518 (\emph{bottom}) and reference spectra for an
L6, T0, and T4. Dotted lines mark the zero points for each
spectrum.}\label{fig:spec}
\end{figure*}

2M~0518's location on the color-color diagram is shown in Figure
\ref{fig:color} and a finder chart is given in Figure \ref{fig:finder}.
 Near-infrared spectroscopy was obtained on 2003 September 19 with SpeX
\citep{spex} on the NASA Infrared Telescope Facility (IRTF) on Mauna Kea.
Observations were taken in low-resolution (R=250) prism mode yielding a
single order from 0.8--2.5~$\micron$, encompassing the $J$-, $H$-, and
$K$-bands (centered at 1.2, 1.6, 2.2~$\micron$, respectively). For the
flux-calibration, an A0 star (HD 36965) was observed immediately after the
target observation and at a similar airmass. This was followed by
acquisition of flat-field and arc-line calibration frames. Conditions were
good, although the data were obtained during morning twilight and at an
airmass of 1.6. The data were flat-fielded, extracted,
wavelength-calibrated, and telluric-corrected with Spextool
\citep{spextool,xtellcor}.   The spectrum of 2M~0518 is shown in Figure
\ref{fig:spec} with reference spectra of an L6, T0, and T4 obtained with
the same instrumental setup. The near-infrared observational properties
for all of these objects are listed in Table \ref{tab:data}.

\begin{deluxetable*}{llcllllllc}
\tablecaption{Near-infrared Observational Properties for Objects Plotted
in Figure~\ref{fig:spec}\label{tab:data}}
\tablehead{\colhead{$\alpha_{J2000}$} & \colhead{$\delta$} &\colhead{Sp.}
& \colhead{2MASS $J$}& \colhead{2MASS $H$ }& \colhead{2MASS $K_s$}&
\colhead{($J-H$)}& \colhead{($H-K_s$)}& \colhead{($J-K_s$)} & \colhead{Ref.}\\
\colhead{} & \colhead{} & \colhead{Type} & & & & & & } \startdata
01 03 32.03 & $+$19 35 36.1 & L6 & 16.29$\pm$0.08& 14.90$\pm$0.06 & 14.15$\pm$0.06 & \phn1.39$\pm$0.10  & \phn0.75$\pm$0.08  & \phn2.14$\pm$0.10 & 1\\
04 23 48.58 & $-$04 14 03.5 & T0 & 14.47$\pm$0.03& 13.46$\pm$0.04 & 12.93$\pm$0.03 & \phn1.00$\pm$0.04  & \phn0.53$\pm$0.05  & \phn1.54$\pm$0.04 & 2,3,4\\
22 54 18.92 & $+$31 23 49.8 & T4 & 15.26$\pm$0.05& 15.02$\pm$0.08 & 14.90$\pm$0.15 & \phn0.24$\pm$0.09  & \phn0.12$\pm$0.17  & \phn0.36$\pm$0.15 & 5,6\\
05 18 59.95 & $-$28 28 37.2 & ?  & 15.98$\pm$0.10& 14.83$\pm$0.07 & 14.16$\pm$0.07 & \phn1.15$\pm$0.12  & \phn0.67$\pm$0.10  & \phn1.82$\pm$0.12 & \\
\enddata
\tablecomments{Units of right ascension are hours, minutes, seconds, and
units of declination are degrees, arcminutes, and arcseconds.}

\tablerefs{(1) \citet{67}; (2) \citet{Geballe}; (3) \citet{SDSS2}; (4)
J.~D.~Kirkpatrick et al. (in preparation); (5) \citet{B02}; (6)
\citet{B04} }
\end{deluxetable*}

\section{Discussion}\label{sec:discussion}

While 2M~0518 lies at the red end of the L dwarf sequence, as shown in
Figure \ref{fig:color}, its spectrum clearly shows methane absorption, the
hallmark characteristic of T dwarfs. Additionally, the relative strengths
of the absorption bands (specifically H$_2$O, CH$_4$, and CO) are
anomalous and are not consistent with either a classical L or T dwarf. The
strong methane absorption in $J$- and $H$-bands, and water absorption in
$J$-band is consistent with a mid-T dwarf (cf. 2M~2254 in Figure
\ref{fig:spec}), whereas the weak methane feature in $K$-band more
resembles an early-T dwarf (cf. SDSS~0423 in Figure \ref{fig:spec}). In
addition, carbon monoxide absorption in the $K$-band is typical of late-L
dwarfs (cf. 2M~0103 in Figure \ref{fig:spec}). In the following section we
consider four possible explanations for the unusual properties of 2M~0518.

\subsection{Possible Scenarios}

\textbf{Youth.} Two characteristics of young objects are reddening by dust
and spectral features indicative of low gravity.  The above-average red
color of young objects is attributed to circumstellar dust or
line-of-sight reddening.  Dust preferentially absorbs shorter wavelength
radiation, changing the slope of the spectrum and reddening the colors,
but cannot account for the anomalous absorption bands strengths observed
in 2M~0518. Low gravity, the other characteristic of youth, does affect
$K$-band spectral features.  In particular, a key signature of low gravity
is a decrease in the flux suppression in $K$-band due to the weakening of
H$_2$ collision-induced absorption (CIA) \citep{lowg,saumon}. However, the
downward slope of the $K$-band peak in the spectrum of 2M~0518 does not
indicate weak CIA H$_2$ absorption.  The unusual spectral features of
2M~0518 cannot be explained by either reddening or low gravity and thus
the object is not likely to be a young T dwarf.

\textbf{Single L/T Transition Object.}  An example of a transition object
is \objectname{SDSS~04234858$-$0414035} which is typed as L7.5 in the
optical and T0 in the near-infrared and also has red colors
(J.~D.~Kirkpatrick et al., in preparation; \citet{Geballe}; Table
\ref{tab:data}). This object is discussed in detail by J.~D.~Kirkpatrick
et al. (in preparation) and its spectrum is shown in Figure
\ref{fig:spec}. In general, L/T transition objects have weak methane
absorption in $K$-band and almost non-existent methane features in $J$-
and $H$-bands \citep{Geballe}. In 2M~0518, we see the opposite---strong
methane absorption in $H$-band while weak at $K$-band. The spectrum of
2M~0518 does not fit the description of a single L/T transition object.

\textbf{Low Metallicity.} The weakness of the $K$-band methane and carbon
monoxide features in the spectrum of 2M~0518 may be attributed to low
metallicity.  In metal poor dwarfs, increased CIA H$_2$ absorption
significantly masks these features and results in blue near-infrared
colors \citep{sdL}.  There is no evidence for enhanced CIA H$_2$
absorption in the spectrum of 2M~0518 and its near-infrared colors are
red, not blue.  It is highly unlikely that 2M~0518 is metal poor.

\textbf{Unresolved L/T Binary.} Late-L and mid-T dwarfs have similar
$J$-band absolute magnitudes but very different near-infrared colors.
Since there is no methane absorption in L dwarfs, they are much brighter
than T dwarfs in $K$-band, and thus their near-infrared colors are also
redder.  For this reason, in a late-L/mid-T binary system, the L dwarf
would be expected to dominate the joint flux distribution longward of
$\sim$1.6~$\micron$, partially filling in the methane absorption in $H$-
and $K$-bands and producing red ($J-K$) composite colors.  This is in
qualitative agreement with the available 2MASS photometry and our spectrum
of 2M~0518.

Of these four options, the last offers the most plausible means of
explaining the observed properties of 2M~0518 and, in the following
section, we discuss the binary scenario in detail.

\subsection{2M~0518 as an Unresolved L/T Binary}

\begin{figure*}[t]
\centering
\includegraphics[width=5in]{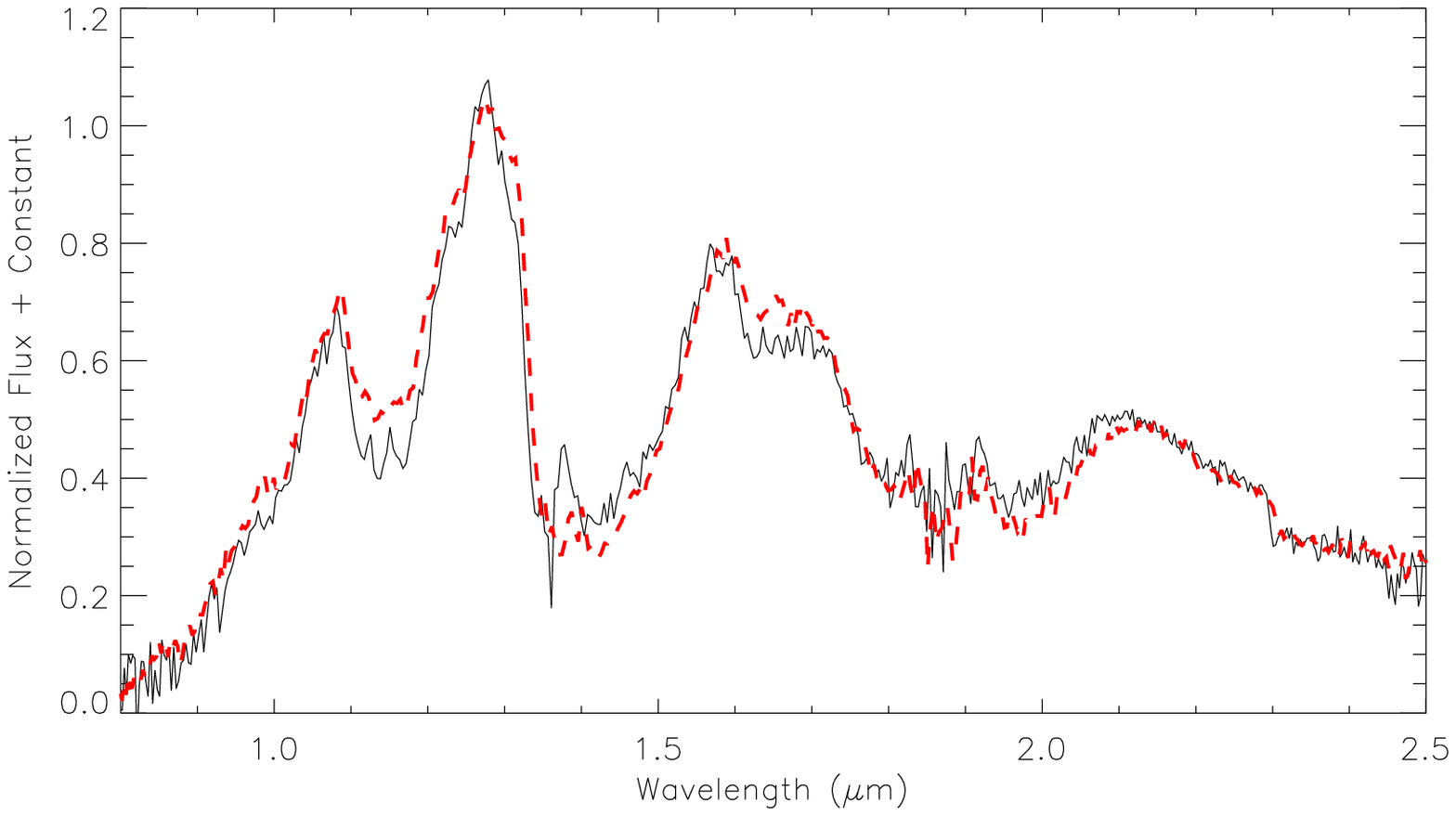}
\caption{Spectrum of 2M~0518 (\emph{solid}) with the scaled sum of 2M~0103
(L6) and 2M~2254 (T4) (\emph{dashed}) superimposed.}\label{fig:zoom}
\end{figure*}


We have made a qualitative attempt at reproducing the spectrum of 2M~0518
by separately combining spectra of four mid/late-L dwarfs with three
early/mid-T dwarfs using various scale factors.  The best qualitative
match is clearly obtained by summing the red L6,
\objectname{2MASS~J01033203$+$1935361}, with the T4,
\objectname{2MASS~J22541892$+$3123498}, weighted by a factor of 1.2, after
both spectra have been normalized to 1.3~$\micron$. The individual spectra
of 2M~0103 and 2M~2254 are shown in Figure \ref{fig:spec} and their scaled
sum is superimposed on the spectrum of 2M~0518 in Figure \ref{fig:zoom}.
The measured colors of the composite spectrum are $(J-H)=0.9$,
$(H-K_s)=0.6$, $(J-K_s)=1.5$ and are comparable to the colors of 2M~0518
(listed in Table~\ref{tab:data}). Based on the overall agreement between
the spectral features and the resultant colors of the combined spectrum
with those of 2M~0518, we find this scenario to be the most compelling and
conclude that 2M~0518 is likely to be an unresolved binary system composed
of a late-L dwarf and a mid-T dwarf.

The two other candidate L/T binary systems,
\objectname{2MASSJ~08503593$+$1057156} and
\objectname{2MASSJ~17281150+3948593}, were discovered with HST WFPC2
imaging \citep{R01,G03}.  Neither these data nor ground based
spectroscopy, however, are able to robustly obtain spectral type estimates
for the secondaries.  J.~E.~Gizis et al. (in preparation) have obtained
HST NICMOS imaging that measures the near-infrared properties of the
individual components and thus will definitively determine if these
systems are comprised of an L and T dwarf or two L dwarfs.

We adopt $M_J=13.9$ for the L6 component using the spectral type/absolute
magnitude relation found by \citet{Tinney03}.  While scaling the T4
component by 1.2 causes the $J$-band peak height of the T4 to be higher
than that of the L dwarf, the T4 still has a fainter $J$ magnitude due to
strong water and methane absorbtion.  We measure $\Delta J = 0.1$,
yielding $M_J=14.0$ for the T4 component.  This estimate is very similar
to the $M_J=13.9$ measured for the T3,
\objectname{SDSS~J10210969$-$0304201}, and $M_J=13.8$ for the T5,
\objectname{2MASS~J05591914$-$1404488}, the two objects with parallax
measurements that have spectral types closest to T4.

While tentative, this observation further supports the argument that there
is a \emph{physical} brightening of T dwarfs at $J$-band compared to L
dwarfs as observed by \citet{Dahn02} and \citet{Tinney03} and is more
likely due to the clearing of clouds and an increased optical depth as
proposed by \citet{cloud} rather than the age selection effects as
suggested by \citet{Tsuji03}.

Combining the individual absolute magnitude estimates for the two
components yields a photometric distance of 36 pc for 2M~0518. If the
2M~0518 system can be resolved and the components can be studied
separately, this object will provide strong constraints on substellar
evolutionary models and will be an important probe of the poorly
understood L/T transition.


\acknowledgments

We would like to thank our IRTF telescope operators, Bill Golish, Dave
Griep, and Eric Volguardsen.  We would also like to thank the referee for
excellent suggestions that improved the manuscript.  This research was
partially supported by a grant from the NASA/NSF NStars initiative,
administered by JPL, Pasadena, CA. K.~L.~C. acknowledges support from a
NSF Graduate Research Fellowship. A.~J.~B. acknowledges support provided
by NASA through Hubble Fellowship grant HST-HF-01137.01 awarded by Space
Telescope Science Institute which is operated by AURA, under NASA contract
NAS5-26555.  This publication makes use of data products from the Two
Micron All Sky Survey, which is a joint project of the University of
Massachusetts and IPAC/CalTech, funded by NASA and the NSF and the
NASA/IPAC Infrared Science Archive, which is operated by JPL/CalTech,
under contract with NASA.

\end{document}